\documentclass[sigchi]{acmart}
\newcommand{\rev}[1]{{\leavevmode\color{black}#1}}%
\copyrightyear{2026}
\acmYear{2026}
\setcopyright{cc}
\setcctype{by}
\acmConference[PerDis '26]{International Conference on Pervasive Displays}{March 16--18, 2026}{Munich, Germany}
\acmBooktitle{International Conference on Pervasive Displays (PerDis '26), March 16--18, 2026, Munich, Germany}
\acmPrice{}
\acmDOI{10.1145/3797993.3798009}
\acmISBN{979-8-4007-2513-5/2026/03}

\usepackage{multirow}%
\usepackage{tabularx}%
\usepackage{color}%

\begin{document}

\title[StatCounter]{%
StatCounter:
A Longitudinal Study of a \\
Portable Scholarly Metric Display%
}%

\author{Jonas Oppenlaender}
\email{jonas.oppenlaender@oulu.fi}
\affiliation{%
  \institution{University of Oulu}
  \city{Oulu}
  \country{Finland}
}%


\begin{abstract}
This study explores 
a handheld, battery-operated e-ink device displaying Google Scholar citation statistics. 
The \textit{StatCounter} 
places academic metrics into the flow of daily life rather than a desktop context.
The work draws on a first-person, longitudinal auto-ethnographic inquiry examining how constant access to scholarly metrics influences motivation, attention, reflection, and emotional responses across work and non-work settings.
The ambient proximity 
and pervasive availability of scholarly metrics invites
frequent micro-checks,
short reflective pauses,
but also introduces moments of 
    second-guessing when numbers drop or stagnate.
Carrying the device prompts new narratives about academic identity, including a sense of companionship during travel and periods away from the office.
Over time, the presence of the device turns metrics from an occasional reference into an ambient background of scholarly life.
The study contributes insight into how situated, embodied access to academic metrics reshapes their meaning, and frames opportunities for designing tools that engage with scholarly evaluation in 
reflective ways.%
\end{abstract}%
%
%
\ccsdesc[500]{Human-centered computing~Displays and imagers}
\ccsdesc[300]{Human-centered computing~Human computer interaction (HCI)}
\ccsdesc[300]{Human-centered computing~Field studies}
%
%
\keywords{
personal informatics, 
self-tracking, quantified self, 
portable display}%
%
%
\begin{teaserfigure}%
\centering
  \includegraphics[width=.89\textwidth]{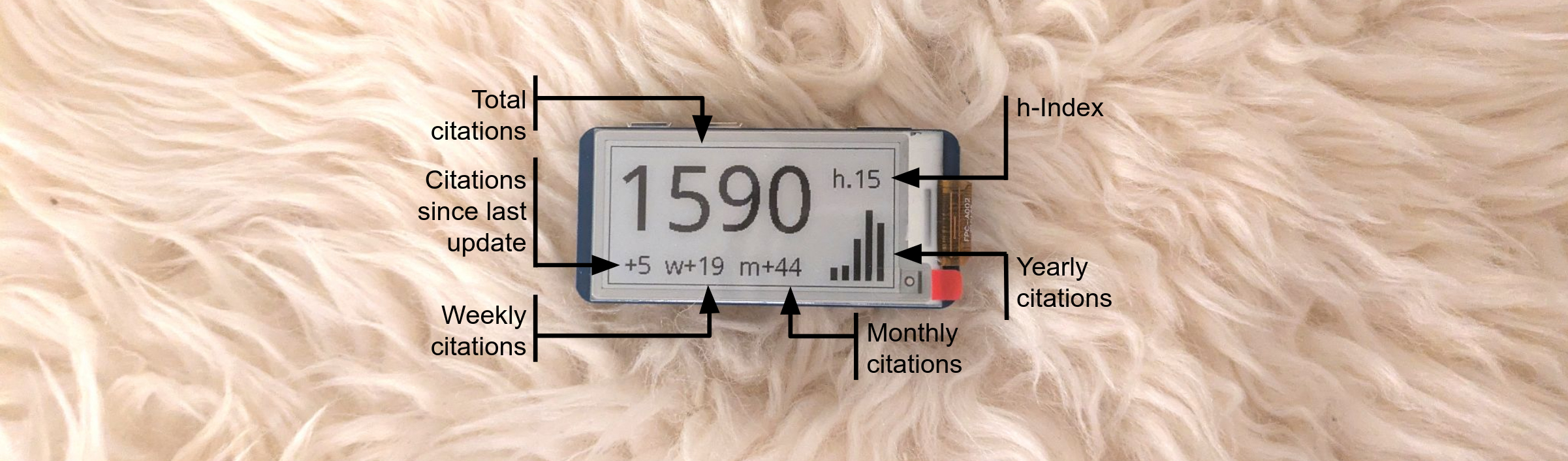}%
  \caption{StatCounter is a handheld e-ink display with scholarly citation statistics collected from Google Scholar.}%
  \Description{A photo of the e-ink display, annotated with labels explaining the scholarly metrics visible: total citations, citations since last update, weekly, monthly, and yearly citations, as well as h-index.}%
  \label{fig:teaser}%
\end{teaserfigure}%
%
\maketitle%
%
%
\section{Introduction}%
%
Academic metrics play a key role in how scholarship is evaluated and discussed.
Prior work has examined how scholarly metrics influence identity, labor, and motivation in academia, touching on work in 
quantified-self practices \cite{big.2012.0002.pdf},
personal informatics \cite{10.1145/3432231},
and how knowledge workers conceptualize personal productivity \cite{paper615.pdf}.
Yet, prior systems often position scholarly metrics as distant indicators accessed through portals in the browser or automated alerts.
I became interested in what happens when citation data becomes
`physicalized' \cite{CHI15_OpportunitiesChallengesDataPhys.pdf}
and ambient \cite{258549.258715.pdf}
as an 
embodied companion rather than an occasional destination on a screen.

To this end, I designed and implemented a small portable device with e-ink display showcasing my Google Scholar citation statistics (see \autoref{fig:teaser}).
The device comfortably fits in a pocket and connects to open wireless networks.
I report on my experiences of carrying the device across workdays, commutes, international conference visits, and moments outside of work.
I~wanted to understand how this 
portable device changes my relationship to scholarly metrics, how it affects attention and emotion, and how it integrates into the background of academic life.

\section{Related Work}%
\label{sec:relatedwork}%
%
%
Researchers use online research platforms, such as Google Scholar\footnote{https://scholar.google.com/}, for self-monitoring their academic metrics \cite{rvab043,Rahwan}.
The scholarly metrics on these platforms, including citation counts and h-index, are widely accepted 
as indicators of academic achievement and progress~\cite{Ma20161}.
Google Scholar is the key source consulted when   evaluating  scientists  for  recruitment  or  promotion purposes \cite{Rahwan}.
While academics routinely track their metrics through institutional databases, there is surprisingly little empirical research studying how academics interact with scholarly data and how they personally self-track their scholarly productivity and impact metrics as a deliberate practice.




\citeauthor{quant-academic}
studied techniques of self-quantification in academia by demonstrating how social networking services enact research and scholarly communication as a 'game' \cite{quant-academic}.
Gamified engagement with the academic ecosystem fulfills a need to be visible in the networked academic world, and
metrics on academic platforms provide a level of legitimacy \cite{rvab043}, but also a way to track progress and performance.
\citeauthor{paper615.pdf} presented a diary study on productivity tracking in knowledge workers \cite{paper615.pdf}, examining how people conceptualize and track personal productivity. However, this study was focused more broadly rather than specifically on academic metrics.
%
%
%
\citeauthor{quantifiedscholar} discuss the growing phenomenon of the ``quantified scholar'' where individual-level impact metrics are increasingly used to measure academic performance, including h-index, downloads, and citations tracked on a daily basis \cite{quantifiedscholar}.
%
Their work provides an overview of metrics used for measuring scholarly productivity, 
discussing how academics track these metrics for tenure, promotion, and grant applications.
\rev{What remains largely unexplored is how scholarly metrics behave when they leave the browser and become a physical, portable presence in everyday life.}%

\begin{figure*}[!thb]%
\centering
\includegraphics[width=.42\linewidth]{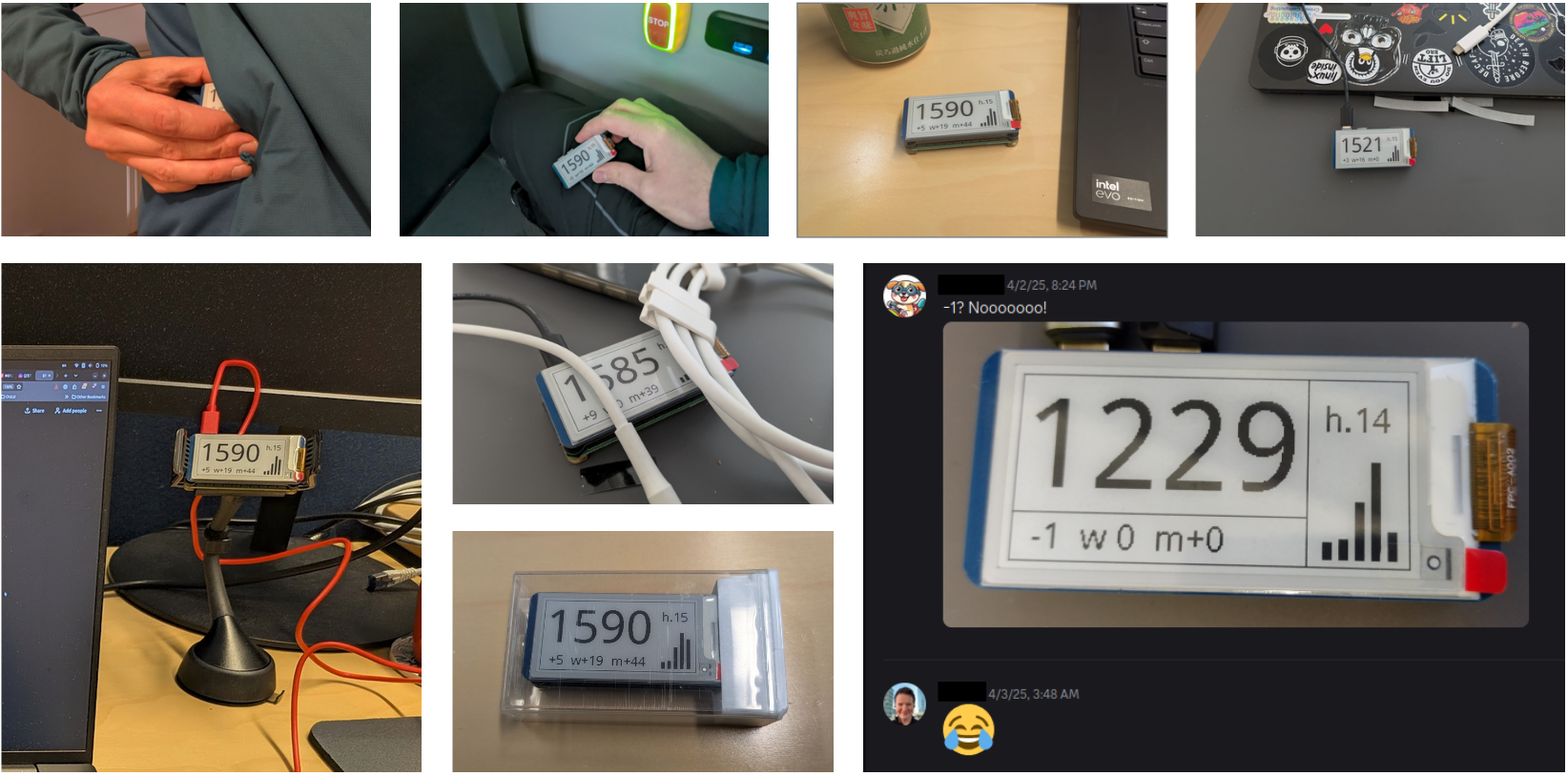}%
\hspace{.3cm}
\includegraphics[width=.34\linewidth]{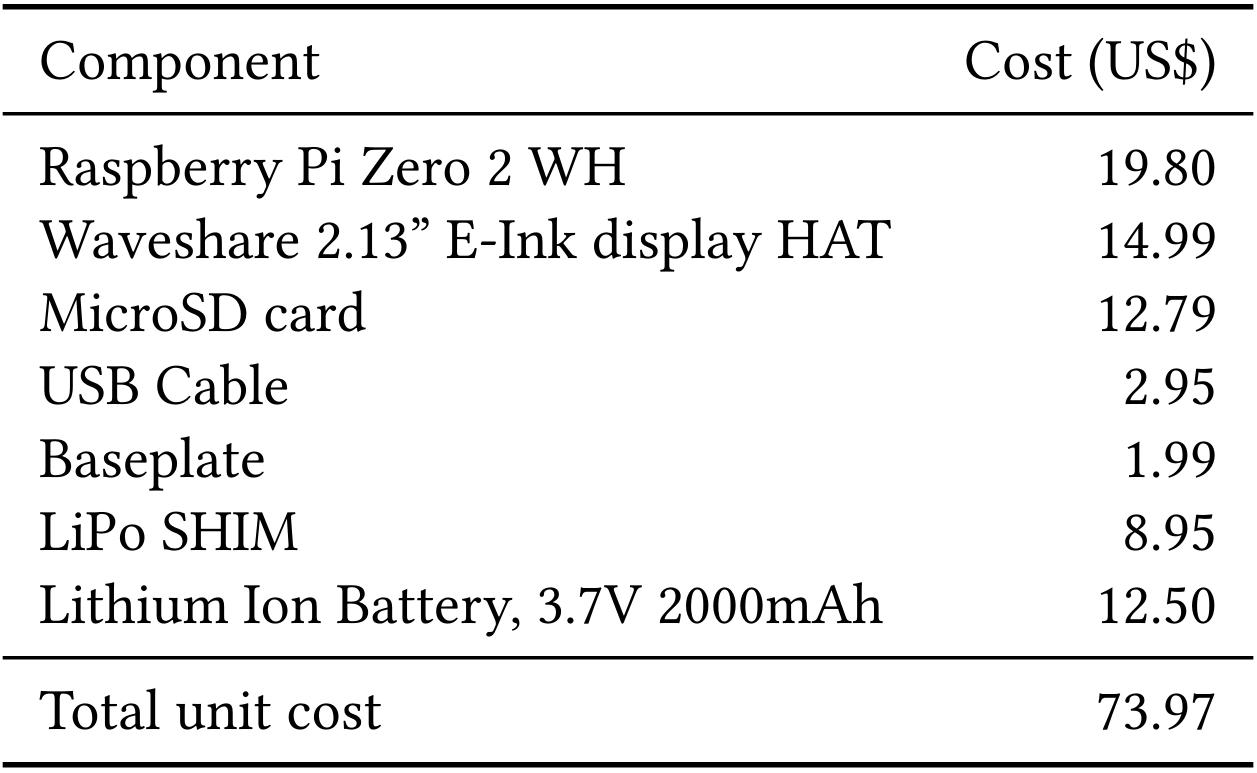}%
\caption{The StatCounter device depicted in 
every-day and work situations during the study (left) and cost breakdown (right)}
\Description{Display design iterations and cost breakdown}
\label{fig:design-iterations}%
\label{tab:cost}%
\label{fig:discord}%
\end{figure*}%

\section{Implementation and Study}%
\label{sec:design}%
%
The StatCounter (see \autoref{fig:teaser}) is a WiFi-enabled Raspberry Zero 2 WH microcontroller board with a Waveshare 2.3'' e-ink display (250x122 pixels resolution), mounted to the Pi's headers.
A bottom plate allows to 
attach a 
battery with double-sided tape, connected to a LiPo adapter which is shimmed between the Raspberry Pi and the display.
The table in \autoref{tab:cost} provides an approximate cost breakdown of the device's components.
The display is driven by a Micropython script that periodically collects citation data from my Google Scholar profile and updates the screen if there is new data (see \href{https://github.com/joetm/raspberry-scholar}{https://github.com/joetm/raspberry-scholar}).
This script is scheduled via a cron job on the device, executing four times per hour between 7 AM and 22 PM.
The e-ink display flashes as a silent visual notification of an update.
%
%
%
%
The form factor of the device allows pocket-based carry across daily movements (see \autoref{fig:design-iterations}).
Connection with the internet is established via a scan of openly accessible wireless networks when on-the-go, with hardcoded network credentials as fallback.
During longer stationary work, the device was docked and kept powered via USB to recharge the battery.

\label{sec:method}%
This study explores how a small, portable display with scholarly information augments the academic life.
I use auto-ethnographic research \cite{4128c4d7-7e05-37a7-afa4-dc62cb32c166,160940690400300403.pdf,3613904.3642355.pdf} to document my experience with the portable device across nine months (March–Nov 2025).
%
During the 36-week deployment period, I carried the device throughout daily life, including commutes, on-campus workdays, a conference trip to Japan, remote work, home routines, and other travels in Europe.
The extended duration allowed shifts in novelty, habituation and interpretation to surface, which aligns with prior auto-ethnographic work that foregrounds temporal progression and longitudinal inquiry \cite{Lucero2021}.
\rev{I used event-driven note-taking~\cite{Larson2014} to record spontaneous reactions to metric changes, producing time-stamped field notes with contextual details and photos. I wrote short analytic memos each week to consolidate early impressions. The corpus was analysed through reflexive thematic analysis~\cite{Braun08082019}, by open coding entries with attention to routines, affective responses, and interpretations of metric changes, then iterated through cycles of coding and 
	consolidation.}%
%
%
%
%
%
%
%
%
\section{Results}%
\label{sec:results}%
\textbf{Micro-checks and the pull of ambient data.}
%
Prior to having the device, I accessed citation data through a browser.
I found myself frequently opening the browser several times a day to see if there had been a change on my Google Scholar profile.
StatCounter made these frequent checks more efficient and ambient: instead of opening a browser window, I now only had to throw a quick glance at the device which always rested in my peripheral vision when working.
The device invited these checks not through explicit (push) notifications but through its ambient presence (and magnetic pull) in my field of vision when flashing with updates. 
When caught in the moment, the flashing update process created an experience of suspense: how much would the display increase?
%
%
%
%
These micro-checking behaviors show how constant availability draws citation data into small openings throughout the day.
The checks were not deliberate actions to monitor scholarly impact; they resembled a subtle gravitational pull exerted by the possibility of change.
This pattern illustrates how ambient data reshapes the cadence of attention, even when the information displayed remains stable.%
%
%
%
%

\textbf{Emotional responses to metric fluctuations.}
The device generated emotional responses  
to both increases and stagnation in metrics.
The emotional dimension of citation counts, normally muted by the infrequency of deliberate browser visits, became an ongoing undercurrent of the day.
Due to being designed to have a rest period over night, the first update of the day was an event to look forward to.
Negative emotional spikes occurred when metrics dropped, such as total citations. 
Fluctuations are normal in citation data, but often go unnoticed.
With StatCounter, however, a drop was immediately visible.
I recall glancing at the display after returning from lunch and noticing that my citation count had decreased by one.
I shared my frustration over this decrease in our research group's social network (see \autoref{fig:discord}).
%
%
%
%
%
Across contexts, the device exposed emotions that normally remain buried in the implicit relationship between researchers and scholarly metrics.
The ambient presence of the scholarly metrics made these feelings accessible, recurring, and tied to the rhythms of daily life.
Through this, the device highlighted how citation metrics function not only as indicators of scholarly impact but also as emotional triggers that shape the texture of academic work life.%

\textbf{Narratives of progress, stagnation, and self-evaluation.}
%
The constant presence of the device shaped how I constructed narratives about progress and scholarly identity. Instead of encountering citation numbers as occasional summaries, I met them as ongoing signals that threaded through the week. These signals prompted small interpretive stories that accumulated into broader stories about how my work was unfolding.
%
%
%
Stagnation also prompted narratives.
During a long stretch in midsummer, 
I started doubting if my personal year-end goals could be reached.
This kind of coupling transformed the metrics into a reflective surface for self-evaluation, even when the relationship between the numbers and my activities was indirect and often felt distant.

Across the 36-week period, narrative episodes showed how ambient metrics cultivate a continuous interpretive stance.
Rather than producing simple reactions, the device invited a sequence of small meaning-making acts that linked numbers with identity, work, and time.
These narratives illustrate how citation counts become part of personal storytelling when they live alongside the rhythms of everyday life.%
%
%

\textbf{The device as a companion across spaces.}
The device took on a companion-like role that traveled with me across environments, modes of activity, and shifts in professional and personal focus.
%
%
For instance, companionship surfaced during travel.
During a multi-day trip for a conference, the device stayed in the outer pocket of my backpack.
%
Companionship was most pronounced during liminal moments: waiting for a bus, brewing coffee in the morning, or grabbing the device when heading to work.
%
%
The companion-like relationship did not arise from the content of the data alone. It stemmed from the device’s materiality, its portability, and its quiet persistence across places where academic metrics rarely appear.
\textbf{Social encounters and the visibility of scholarly metrics.}
Although the StatCounter primarily addressed my own practice, it was also visible to others and became a topic of conversation in several settings.
Reactions ranged from curiosity and amusement to discomfort, and these social responses fed back into how I positioned and interpreted the device.
Colleagues first noticed the StatCounter during informal meetings in my office. The device sat near my laptop. A visitor paused mid-sentence, leaned forward, and asked, ``Is that your citation count?''
The question turned into a brief discussion about whether the device functioned as a tool, a joke, or a status display.
The encounter made me aware that the same numbers that felt companion-like to me could appear as self-promotion or pressure to others.
%
%
Outside academic contexts, reactions 
foregrounded the oddness of carrying such a display.
During a train journey, a stranger noticed the device in my hand and asked whether it was ``a kind of pager.'' When I explained that it showed citations, the person responded with a mixture of confusion and curiosity: ``You carry around how much people cite you?''
The conversation exposed how naturalized academic metrics seem within academia and how unusual they appear elsewhere.
%
%
Across these encounters, the StatCounter operated as a social object that exposed the presence of metrics in contexts where they are usually implicit or absent.
People's reactions made visible the norms and tensions around academic evaluation, self-tracking, and status.
These interactions showed that ambient scholarly displays do not only shape individual experience, they also participate in social negotiations around what kinds of academic data are acceptable to show, share, or hide in everyday life.
\rev{Taken together, these patterns show that a portable scholarly display does not merely relocate citation data. It introduces new routines, emotional textures, and social negotiations that reframe how metrics participate in academic life.}%
%
%
%
\section{Discussion and Conclusion}%
\label{sec:discussion}%
%
The continuous access to embodied citation data shaped my daily reflective academic practice across work and non-work settings,
and the handheld metric display became an ambient companion.
\rev{That citation counts provoke emotional responses is, on its own, unsurprising. What this study surfaces is how the material shift from browser to pocket changes the character of that relationship. When metrics travel into domestic, social, and transit spaces, they become a companion that participates in contexts where academic evaluation is normally absent. The same ambient access that supports reflection also extends evaluative pressure into spaces that might otherwise offer distance from it. This tension between reflective benefit and evaluative reach is, I believe, the central design challenge for ambient scholarly displays.}
In this work, I provided a set of design considerations for portable systems supporting reflective engagement with scholarly metrics.
Together, these directions outline a research agenda for understanding and designing ambient encounters with embodied scholarly metrics, with attention to
how such encounters can augment
    the temporal, material, and affective dimensions of 
    academic life.

Future work could investigate not just displaying personal information, but aggregation of scholarly metrics for research groups or whole departments.
Future studies could also include researchers from different disciplinary communities, career stages, and institutional environments to use comparable devices. This would broaden the perspective beyond the single-user account presented in this article.
Adopting a multi-participant lens would support comparative analyses of how local cultures of evaluation, publication expectations, and personal orientations toward metrics shape engagement with scholarly metrics and data.
Such work could also reveal tensions that emerge when ambient metric access interacts with workload, work precarity, or collaborative research practices.%

\begin{acks}
The author gratefully acknowledges support from the University of Oulu’s Centre for Applied Computing (CAC) and the NOSTE development programme for education.
\end{acks}

\bibliographystyle{ACM-Reference-Format}
\bibliography{main}


\end{document}